\documentclass[structabstract]{aa}

\usepackage{graphicx} 
\usepackage{colortbl}
\usepackage{xcolor}
\usepackage{url}
\usepackage{booktabs}
\usepackage[small,hang]{caption}
\usepackage{txfonts}
\usepackage{natbib} 
\bibpunct{(}{)}{;}{a}{}{,} 

\begin{document}

\newcommand{\MPIA}{1} 
\newcommand{\bfref}{}

\title{Polarisation of very-low-mass stars and brown dwarfs}
\subtitle{I. VLT/FORS1 optical observations of field ultra-cool dwarfs
\thanks{Based on observations collected at the European Observatory, Paranal, Chile,
             under programmes 075.C-0653(A) and 077.C-0819(A).}
             }
\author{
  B. Goldman\inst{\MPIA}
    \and
  J. Pitann\inst{\MPIA}
    \and
  M.~R. Zapatero Osorio\inst{2} 
    \and \\
 C. A. L. Bailer-Jones\inst{\MPIA} 
    \and
  V. J. S.~B\'ejar\inst{2} 
    \and
  J. A.~Caballero\inst{3}
    \and
  Th.~Henning\inst{\MPIA}
}
   \date{Received 14 October 2008; accepted 15 January 2009}

   \offprints{B.~Goldman, {\tt go{\,\hspace{-1pt}}ld\hspace{-1pt}\,man{\,}@mp{}ia.de}}

   \institute{
              Max Planck Institute for Astronomy, K\"onigstuhl~17, D--69117 Heidelberg, Germany
         \and 
              Instituto de Astrof{\'\i}sica de Canarias, C/ V\'\i a Lact\'ea S/N, E-38205 La~Laguna, Tenerife, Spain
         \and 
              Departamento de Astrof\'{\i}sica, Facultad de Ciencias F{\'\i}sicas, Universidad Complutense de Madrid, E--28040 Madrid, Spain
       }

\abstract
 {Ultra-cool dwarfs of the L~spectral type ($T_{\rm eff}=1400$--2200\,K) are known to have dusty atmospheres. Asymmetries {of the dwarf surface} may arise from rotationally-induced flattening and dust-cloud coverage, and {may} result in non-zero linear polarisation through dust scattering.}
  {We aim to study the heterogeneity of ultra-cool dwarfs' atmospheres and the grain-size effects on the polarisation degree in a sample of nine late M, L and early T dwarfs.}
  {We obtain {linear} polarimetric imaging measurements using FORS1 at the Very Large Telescope, in the Bessel $I$ filter, and for a subset in the Bessel $R$ and the Gunn $z$ filters.}
  {We measure a polarisation degree of $(0.31\pm0.06)$\% for LHS102BC.
We fail to detect linear polarisation in the rest of our sample, with upper-limits on the  polarisation degree of each object of {0.09}\% to {0.76}\% (95\% of confidence level), depending on the targets and the bands.
For those targets we do not find evidence of large-scale cloud {horizontal} structure in our data.
{\bfref Together with previous surveys}, our results {\bfref set} the fraction of ultra-cool dwarfs with detected linear polarisation to $30^{+10}_{-6}$\% (1-$\sigma$ errors). 
From the whole sample of well-measured objects with errors smaller than 0.1\%, the fraction of ultra-cool dwarfs with polarisation degree larger than 0.3\% is smaller than 16\% (95\% confidence level).
}
 {
For three brown dwarfs, our observations indicate polarisation degrees different (at the 3-$\sigma$ level) than previously reported, giving hints of possible variations.
Our results fail to correlate with the {\bfref current model} predictions for ultra-cool dwarf polarisation for a flattening-induced polarisation, or with the variability studies for a polarisation induced by an heterogeneous cloud cover.
This stresses the intricacy of each of those tasks, but may as well proceed from complex and dynamic atmospheric processes.
}

\maketitle

%-------------------------------------------------------------------

\section{Introduction} \label{intro}

The L spectral type covers effective temperatures of 2200\,K to $\approx1400$\,K \citep{Kirkp99,Marti99}.
The lowest-mass stars on the main sequence as well as substellar objects with a large range of masses, at a given time of their cooling sequence, populate this spectral type.
The optical spectra of L~dwarfs show specific features that define the type: 
strong hybrid bands (FeH and CrH) and alkali lines (Na~{\sc i}, K~{\sc i}, Rb~{\sc i}, Cs~{\sc i}).
The oxides TiO and VO produce weak absorption features, which is interpreted to be due {to} the coagulation of those species into dust grains.
Other refractory elements such as Al, Ca, Fe, and Mg condensate into grains and form dust clouds.
By contrast, in the warmer atmosphere of the early M~dwarfs, the temperature is too high for the molecules of e.g. TiO or VO to aggregate into grains, while the atmospheric models show that the clouds made of larger dust grains sediment below the photosphere of the cooler mid-T~brown dwarfs.

Several studies have shown that ultra-cool dwarfs are fast rotating \citep{Mohan03,CBJ04,ZOsor06}. It is therefore likely that the rotation induces a flattening at the poles.
Furthermore, the fast rotation may increase the turbulence in their convective atmospheres, leading to possible heterogeneities in the cloud deck.
{The polarisation arising from many small-scale, randomly-distributed structures would cancel out, but large-scale heterogeneities may result in net polarisation.}
{\bfref Gravitational pulling by a close companion could also break the symmetry of the brown dwarf surface, as it is expected rotational acceleration does. }

\citet{Sengu01} {\bfref calculated the polarisation of rotationally-flattened ultra-cool dwarfs} and predicted that they could have non-zero polarised light. They estimated the polarisation degree due to the rotationally-induced flattening for single scattering and various grain-size distributions, while \citet{Sengu05} studied the effect of multiple scattering.

Later, two observational studies took on to test those predictions.
\citet{Menar02} (M02 hereafter) used the VLT/FORS1 imager in Paranal Observatory to observe in the Bessel $I$~band a sample of eight L0 to L8 ultra-cool dwarfs (and binaries), detecting a significant, {although small} polarisation degree in three of those, {between 0.1\% and} 0.2\%.
\citet{ZOsor05} (ZO05) used the 2.2m/CAFOS imager in Calar Alto Observatory to observe a larger sample of 43~ultra-cool dwarfs with spectral types between M6 and L7.5, in Johnson $R$ and $I$ bands.
They found eleven dwarfs with significant polarisation, including seven with an estimated polarisation degree larger than 0.5\%.

\begin{table*}
\caption[]{Summary of targets and references, ordered by spectral type.}
\begin{tabular}{lccccc}
            \hline
            \hline
            \noalign{\smallskip}
                {Object} & $I$ & $J$ & {Sp.T.} & Discovery paper & $v\sin i$ [km/s]  \\ %
            \noalign{\smallskip}
            \hline
            \noalign{\smallskip}
    DENIS J22000201-3038327        & 16.7~~ & $13.44\pm0.03$  & M9+L0       & \citet{Kenda04}    & 17$^1$ \\ %
     Kelu\,1\,AB                                       & 16.94& $13.41\pm0.03$  & L2          & \citet{Ruiz97}       & 60$^2$ , 76$^1$\\ %
     2MASS J00361617+1821104  %(2MUCD 20029) 
                                                                  & 16.11 & $12.47\pm0.03$  & L3.5        & \citet{Reid00}       & $36\pm2.7^3$, 45$^1$ \\ %
     2MASS J21580457$-$1550098 %
                                                                  & 18.5$^7$ & $15.04\pm0.04$    & L4          & ...  & ... \\ %
     LHS 102BC (GJ 1001BC)               &16.68 & $13.11\pm0.02$ & L4.5+L4.5$^4$      & \citet{Goldm99}    & 42$^1$ \\ %
     2MASS J15074769$-$1627386 %
                                                                  & 16.65 & $12.83\pm0.03$ & L5          & \citet{Reid00}       & 27.2$^5$, 32$^1$ \\ %
     DENIS-P J225210.7$-$173013       & 17.9~~  & $14.31\pm0.03$ & L6:+T2:$^6$ & \citet{Kenda04}    & ... \\ %
     2MASS J224431.67+204343.3                                 & 20.4$^7$ & $16.48\pm0.14$ & L7.5        & \citet{Dahn02}       & ... \\ %
     $\epsilon$ Indi B                                  & 16.6~~ & $11.91\pm0.02$ & T1+T6       & \citet{Schol03}     & ... \\ %
            \noalign{\smallskip}
            \hline
  \end{tabular}
  \begin{tabular}{p{15cm}}
$^1$~\citet{Reine08}; $^2$~\citet{Mohan03}; $^3$~\citet{ZOsor06}; $^4$~\citet{Golim04bi}; 
$^5$~\citet{CBJ04}; $^6$~\citet{Reid06bi}; $^7$ Estimates based on ultra-cool dwarfs of similar spectral type.
  \end{tabular}
  \label{tab:Objects}
\end{table*}

Given its dependency on the rotationally-induced flattening, the grain-size distribution and the {large-scale} cloud distribution on the dwarf surface, polarimetric measurements have the potential to shed light on many fundamental processes in ultra-cool atmospheres, which are currently hardly constrained by other observing modes, or not at all. 
At the same time, the {entangled dependency} of the polarisation degree on those many parameters requires a specific observing strategy. 
Rotational velocities could be correlated to the polarisation degree.
Dynamical weather-like cloud patterns could be detected through polarimetric monitoring.
Finally, the dependence of the polarisation degree to the grain size can be probed through multi-wavelength observations. 
Comparison of optical and near-infrared data, particularly if the typical grain size is in the 1-$\mu$m range, is promising.

This program constitutes an ambitious and demanding observing challenge. In this article we report on the first, likely easiest stage, namely polarimetric observations in the red {\bfref part of the spectrum}, {in one to three filters depending on the targets}. As some targets were previously observed by M02 or ZO05, {\bfref multiple-epoch observations} are already available.
In Section\,\ref{observations}, we describe our sample and the observations, 
as well as our data reduction and polarisation determination.
We present our results in Section\,\ref{results}, and we compare and combine them with previous measurements.
Finally we give our conclusions in Section\,\ref{conclusion}.

\section{Observations} \label{observations}

\subsection{Target selection}  \label{selection}

We searched the compendium DwarfArchives\footnote{\tt http://www.DwarfArchives.org} and chose our targets according to the following criteria:
\begin{enumerate}
\item Availability of published polarimetric measurements, in order to search for long-term variability.
\item Spectral type: we selected targets sampling a large range of spectral types from M9 to early T.
\item Brightness: in order to achieve signal-to-noise ratios (SNR) of 500 and more in each image in a few minute exposure times, in a wide range of observing conditions, we selected the brightest targets in each spectral type.
\item {If possible,} availability of high-resolution spectroscopic determination of the rotation velocity.
\end {enumerate}
The targets are described in Table\,\ref{tab:Objects}\footnote{We used the Ultracool Dwarf Catalog hosted at {\tt http://www.iac.es/galeria/ege/catalogo\_espectral/} to obtain the  $I$ magnitude or estimate it from ultra-cool dwarfs of similar spectral type.}.

\subsection{Observing conditions and instrumental set-up}

All data were obtained using the Focal Reducer and low dispersion Spectrograph FORS1 \citep{Appen98}, which is mounted at the Cassegrain focus of the UT2 (Kueyen) of the VLT facility.
With some exceptions, we used the $\rm 2k\times 2k$ Tektronix CCD and four output amplifiers. In some cases the CCD was windowed to the central $\rm 1k\times 1k$ area.
In both cases no binning was applied and the pixel scale is 0\farcs2, and the full field of view is $6.8\times6.8\,\rm min^2$.
The {\sc ipol} polarimetric imaging mode of FORS1 uses a Wollaston prism splitting the beam into the ordinary and extraordinary parts; a rotating half-wave retarder plate; and a 11\arcsec-wide strip mask to prevent overlapping of the two beams.
All the targets were centred at the same CCD position, in the middle of the central strip.
Observations were carried out through the Bessel $R$ and $I$ filters, as well as the Gunn $z$ filter, with central wavelength and full-width at half maximum of $657\pm150$\,nm, $768\pm138$\,nm, and $910\pm131$\,nm respectively.

\begin{table}[b]
  \caption{Observing log.}
  \begin{tabular}{lcccccc}
            \hline
            \hline
    \noalign{\smallskip}
    {Target} & {Date } & {Filt.} &{Seeing$^1$} &{SNR$^2$}  & $t_{\rm exp}$ & $N_\theta^3$ \\
                    &              &          &  $['']$       &               & $[\rm s]$ \\
    \noalign{\smallskip}
    \midrule
    \noalign{\smallskip}
    DENIS J2200  	& 01/08/05 &$I$ 	& 1.7 	&  720	& 160	&  3 \\
    \noalign{\smallskip}
    Kelu\,1\,AB   		& 02/08/05 &$I$ 	& 1.4 	& 529	& 200	&  2 \\
    			              & 02/08/05 & z  	& 1.6 	& 243	& 200	&  2 \\
    \noalign{\smallskip}
    2MASS J0036 	&  01/08/05 &$R$	& 1.5 	&706	& 600	&  3 \\
		                   	& 01/08/05 &$I$	& 1.7  & 677	& 150	&  3 \\
							& 10/08/06 &$I$	& 1.3 	 & 785	& 150	&  4 \\
							& 11/08/06 &$I$	& 0.9 	 & 937	& 150	&  4 \\
    \noalign{\smallskip}
    2MASS J2158 	& 01/08/05 &$I$ 	& 1.2 & 189	& 300	&  3 \\
							& 02/08/05 & z 	& 1.0 	 & 350	& 400	&  3 \\
    \noalign{\smallskip}
    LHS\,102BC      	& 01/08/05 &$I$ 	& 1.2 	 & 186	& 130	&  2 \\
                                 & 02/08/05 &$I$ & 0.9 	 & 192  & 180	&  2 \\
							& 02/08/05 & z 	& 0.9 	 & 532	& 200	&  3 \\
							& 28/07/06 & I	& 1.2	 & 890	& 150	&  4 \\
    \noalign{\smallskip}
    2MASS J1507 	& 01/08/05 &$R$	& 1.0 	 & 491	& 200	&  3 \\
							& 01/08/05 &$I$ 	& 1.0 	 & 783	& 160	&  3 \\
    \noalign{\smallskip}
    DENIS J2252 	& 02/08/05 &$I$ 	& 1.0 	 & 428	& 600	&  3 \\
    \noalign{\smallskip}
    2MASS J2244 	& 02/08/05 & z & 1.1  & 124	& 600	&  2 \\
                                 & 19/08/06 &$I$& 0.8 	& 222	& 300 &  3 \\
                                 & 21/08/06 &$I$& 0.9 	 & 214	& 300 &  3 \\
							& 22/08/06 &$I$	& 0.8 & 229	& 300	&  3 \\
    \noalign{\smallskip}
    $\epsilon$ Indi Bab & 01/08/05 & $I$  	& 1.2 	 & 422	& 150	&  3 \\
    \noalign{\smallskip}
    \bottomrule
  \end{tabular}
  \begin{tabular}{p{8.5cm}}
  $^1$ The seeing is calculated from the average FWHM of the science target in the given night. 
 $^2$ {\bfref Signal-to-noise ratio (SNR), averaged for each target over a night}.
  $^3$ Number ($N_\theta$) of exposures per retarder angle. 
  \end{tabular}
  \label{tab:Observations}
\end{table}

We used four retarder angles: $0\deg$, $22.5\deg$, $45\deg$, and $67.5\deg$, which allow to derive the Stokes parameters (see Section\,\ref{polmes}).
All science targets were observed {\bfref several} {consecutive} times at each retarder angle, while the bright polarimetric standard stars were observed only once (see Table\,\ref{tab:Observations}).

The observations were conducted in two programmes, one in visitor mode, 
the second in service mode (see Table\,\ref{tab:Observations}).
The atmospheric transparency was good at all times.
The seeing conditions were {average for the site} during service mode, 
but they were worse than average during the two visitor nights, with a median seeing of 1.5\arcsec, and strong winds from the South-East.

\subsection{Polarimetric measurement and error determination} \label{polmes}

The images were bias-subtracted and flat-fielded with twilight flat-field of the same or contiguous nights, obtained without the Wollaston prism, using the standard {\sc noao/iraf} routines \citep{Tody93}.
Cosmic rays {and isolated bad pixels were removed from the flat-field frames using the {\tt crreject} {\sc iraf} routine, and from the science images} using the {\tt sigma\_filter} {\sc idl} routine.
The flat-fielding is {\bfref an important} issue, as the flat-fielding inaccuracies (at the extraordinary position relative to the ordinary position) will propagate linearly, to first order, into the $Q$ and $U$ measurements.
{\bfref This problem is mitigated in our case because all frames of the targets and standard stars are observed at the same CCD pixels, at an accuracy better than the full width at half maximum (FWHM) of the stellar profile.}
{\bfref The effects of the flat-field photonoise and systematic errors on each pixel's sensitivity} are reduced by the excellent sampling of the point-spread function (PSF). A large number of pixels (typically several dozens) are illuminated {\bfref by the star} and used to measure its flux, so that the {\it non-correlated} effect {\bfref roughly} scales down as the square-root of that number.

{\bfref Correcting for the linear polarisation of usual flat-field sources including twilight sky and internal screens is difficult, so we obtained our flat-field frames without the Wollaston prism.
Therefore they do not account for the whole light path.
In fact, we measure a $\approx 0.7\%$ systematic difference between the ordinary and extraordinary fluxes in all our observations, independent of the filter, observing conditions, integration time or target magnitude.
Regardless of the cause of this systematic difference, it mostly vanishes as for the polarisation determination we take the differences of the relative differences of the ordinary and extraordinary fluxes (see Eqs. below).
}
In addition, the ordinary and extraordinary rays swap positions for certain pairs of rotator positions \citep[when the angles differ by $\pi/4$, see][]{Patat06}.
Ultimately the {\bfref effects of the in}accuracies of the flat-field are estimated using the polarimetric standard stars' observations (see next subsection). 
However the spectral types and spectral energy distributions of the scientific targets and the standard stars greatly differ.

We derive the polarimetric information in a way similar to that of ZO05.
We extract $\rm 30\times 20$-arcsec$^2$ stamp images around the target, {with some adjustments when a nearby star could affect the background measurement}.
The fluxes of the ordinary and extraordinary stamp images are measured with aperture photometry using using {\tt SExtractor} \citep{Berti96}, as well as PSF fitted photometry using the {\sc iraf} {\tt daophot} program.
{For consistency, we chose a constant aperture of 10\,pixels, or 2.0''.}
We then average the (extra)ordinary fluxes of the 2--4~images at a given retarder angle, and use the standard deviation as a proxy for the photometric error per frame.
With these averages, we calculate the normalised Stokes parameters, $Q/I$ and $U/I$ with the formulae \citep{ZOsor05,Patat06}: 
   \begin{eqnarray*}
      F(\eta)  &=&  \frac{f_O(\eta)-f_E(\eta)} {f_O(\eta)+f_E(\eta)} \\
      P_Q = Q/I  &=&  \frac{1}{2} (F(0\degr)-F(45\degr))  \\
      P_U = U/I  &=& \frac{1}{2} (F(22.5\degr)-F(67.5\degr)) \\
      P_L  &=& \sqrt{(Q/I)^2+(U/I)^2} \label{eq:P}, \\
   \end{eqnarray*}
where $f_O(\eta)$ and $f_E(\eta)$ are the ordinary and extraordinary averaged fluxes measured at the retarder angle $\eta$.

\begin{figure}[t]
\includegraphics[width=.5\textwidth]{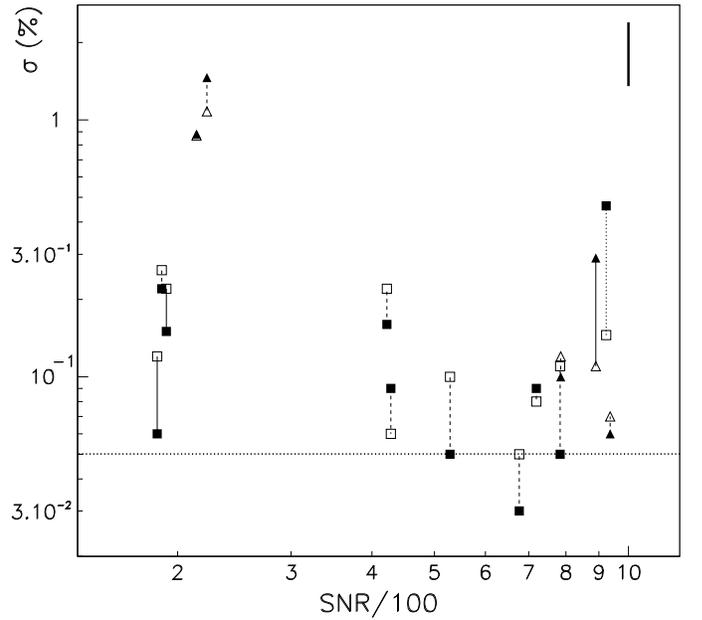}
   \caption{Polarimetric accuracy $\sigma(P_Q)$ (filled symbols) and $\sigma(P_U)$ vs. signal-to-noise ratio {\it per frame} (SNR), in the $I$ band (squares: P75, triangles: P77).
   {\bfref The SNR is the averaged SNR over all extraordinary and ordinary fluxes for the given object.}
   The $\sigma(P_Q)$ and $\sigma(P_U)$ values are linked by a dashed line for a given target, for clarity (solid line for LHS102BC, dotted line for WD1344+106).
   The horizontal line indicates the theoretical limit for $\sigma(P_L)$ of \citet{Patat06}.
     The vertical thick line indicates the statistical 1-$\sigma$ {\bfref uncertainty on the error} expected for three measurements per retarder position.
   }
      \label{errors}
\end{figure}

  \begin{table*}
	\caption{Polarisation degrees and observing dates of standard stars: highly-polarised and (in italics) non-polarised. 
    {\bfref The $Q/I$ and $U/I$ values are {\it not} corrected for the pleochroism (see text).}
	The reference polarisation degrees {\bfref and angles} are from \citet{Fossa07}. 
	}
	\begin{tabular}{{lclcccccc}}
	\toprule
	{Object} & Date (UT) & {Filter} & $Q/I\;[\%]$ & $U/I\;[\%]$ & $P_{\rm Obs}\,[\%]$ &  $P_{\rm Foss}\,[\%]$ & $\theta_{\rm Obs} (^{\rm o})$ & $\theta_{\rm Foss} (^{\rm o}) $ \\
	\midrule 
	{\it WD1615-154} & 02/08/05  & R-Bessel & $+0.13$ & $+0.16$ & $0.14$ &  ... & $-25$ & ...\\
	{\it WD1344+106} & 01/08/05  & I-Bessel & $-0.07\pm0.13$ & $+0.13\pm0.11$ & 0--0.199 &  ... & $+34\pm23$ & ...\\ 
	\midrule 
	{\it WD0310-688} & 19/08/06 & I-Bessel & $+0.17$ & $+0.03$ & $0.17$ & ... & +5   & ... \\
	{\it WD0310-688} & 21/08/06 & I-Bessel & $+0.05$ & $+0.06$ & $0.08$ & ... & +24 & ... \\
	\midrule 
	Hiltner 652            & 03/08/05	& V-Bessel & $+6.29$ & $+0.33$ & 6.30 & $6.25\pm 0.01$ & 179.7 & $179.18\pm0.20$\\
	Hiltner 652           & 03/08/05	  & R-Bessel & $+5.90$ & $-0.28$ & 5.91 & $6.07\pm0.02$ & 179.8& $179.39\pm0.10$\\
	NGC 2024            &03/08/05	  & R-Bessel & $-0.22$ & $-9.63$ & 9.63 & $9.62\pm 0.01$ & 45.5 & $135.84\pm0.02$\\
	NGC 2024            &03/08/05  & I-Bessel & $-0.48$ & $-9.13$ & 9.14 & $9.12\pm 0.04$ & 46.4 & $136.26\pm0.15$\\
	NGC 2024             & 03/08/05	& V-Bessel & $+0.57$ &$-9.58$ & $9.60$ & $9.65\pm 0.06$ & 134.9 & $135.47\pm0.59$\\
	\midrule 
	Hiltner 652            & 01/08/06  & I-Bessel & $+5.43$ & $-0.64$& $5.47$ & $5.61\pm0.04 ^1$& 179.5  & $179.18\pm0.11^1$\\
	Hiltner 652            & 02/09/06  & I-Bessel & $+5.55$ & $-0.59$ & $5.52$ & $5.61\pm0.04^1$& 179.8 & $179.18\pm0.11^1$\\
	NGC 2024            &03/09/06	  & R-Bessel & $-0.11$ & $-9.58$ & 9.58 & $9.62\pm 0.01$ & 45.9 & $135.84\pm0.02$\\
	NGC2024              & 10/08/06	& I-Bessel & $-0.59$ & $-9.16$ & 9.18 & $9.12\pm 0.04$ & 46.0 & $136.26\pm0.15$\\
	BD-125133           & 02/09/06	& I-Bessel & $+1.04$ & $-3.38$ & 3.54 & $3.57\pm0.09$ & 146.4 & $143.99\pm2.27$\\
	\bottomrule
	\end{tabular}
  \begin{list}{}{}
  \item[$^1$]  \citet{Fossa07} only reports measurements using the PMOS mode. For the $B$ and $V$ filters for which they report measurements in both IPOL and PMOS modes, one finds differences up to $(0.10\pm0.02)$\% and $(0.38\pm0.14)\deg$.
  \end{list}
    \label{polstd}
  \end{table*}
  
We propagate the photometric errors in order to obtain the errors on $Q/I$ and $U/I$ (see Fig.\,\ref{errors}). 
Using the standard deviation distribution {of multiple $F(\eta)$ measurements} is a robust, model-independent way to estimate the polarimetric errors. In particular, it is insensitive to errors in the CCD gain or in the noise model. However, it does not include flat-fielding errors, which in our case have a systematic effect for each target.
Furthermore, the small number of images (generally $4\times 3$) results in a large relative uncertainty (roughly 30\%).
{The case of the polarisation degree is more complicated. The error propagation gives:
\begin{eqnarray*}
\sigma(P_L) &=& \sqrt{\frac{P_Q^2.\sigma(P_Q)^2+P_U^2.\sigma(P_U)^2)}{P_Q^2+P_U^2}}
 \end{eqnarray*}
Because it is the quadratic sum of two variables, that are roughly Gaussian distributed, its distribution is biased toward non-zero measurements \citep[see e.g.][ and references therein]{Fendt96,Patat06}. This problem has been studied in a number of publications, and we adopt the correction of the polarisation degree proposed by \citet{Simmo85}, as well as their confidence interval determination---their Fig.6, taking {\bfref as normalised polarisation degree} \mbox{$p=\sqrt{P_Q^2/\sigma(P_Q)^2+P_U^2/\sigma(P_U)^2}$}.

When a significant polarisation degree is measured, we calculate the polarisation angle, which is unbiased \citep{Wardl74}. We follow \citet{Landi07}, whose expression takes care of the usual discontinuity around $P_U=0$:
\begin{eqnarray*}
\theta                &=& {1 \over 2} \, {\rm sign}(P_U) \arccos \left(\frac{P_Q}{\sqrt{P_Q^2 + P_U^2}} \right) \\
\end{eqnarray*}
For significant polarisation degrees \citep[$P_L>3\sigma(P_L)$, ][]{Wardl74}, the bias on the polarisation angle becomes negligible and we simply propagate the errors:
\begin{eqnarray*}
\sigma(\theta) &=& \sqrt{\frac{\sigma(P_U)^2}{|P_Q+P_U^2/P_Q|}+\frac{P_U^2\sigma(P_Q)^2}{(P_Q^2+P_U^2)^2}}
 \end{eqnarray*}
}

FORS1 suffers from instrumental polarisation in the edges \citep{Patat06}, but it is negligible at the centre of the chip.
{We repeatedly observed our targets at this central position, so that our observations do not suffer from instrumental polarisation.}
We measured the polarisation degree of stars in the field of view to check the instrumental polarisation, flat-fielding and interstellar polarisation. For those stars away from the centre, we corrected the polarisation degree using the data of \citet{Patat06} in the $I$ band. No such data are available in the $z$ band.

\subsection{Polarimetric standard stars} \label{std}

FORS1 has proved to be extremely stable in its polarimetric mode \citep[M02; ][]{Patat06}.
The instrumental polarisation is checked on a monthly basis with highly polarised standard star observations, and more frequently with non-polarised standard stars.
{\bfref It appears that the polarisation angle needs to be corrected for instrumental polarisation by a constant (filter-dependent) amount,\footnote{\bfref The effect is called pleochroism \citep{Patat06}. For this chromatic zero angle, we used the values indicated in the FORS1+2 manual, section 4.6.2: $-1.19\deg$ in $R$, $-2.89\deg$ in $I$, $-1.64\deg$ in $z$.} while no correction in the polarisation degree is required.}

Our error determination procedure requires multiple images for each retarder angle, while we only have one for the bright standard stars.
{\bfref \citet{Patat06} provide an estimate of the minimal uncertainties that can be obtained with the four-retarder-position procedure of  0.05\% (see Fig.\,\ref{errors}).
Six of our polarimetric measurements of polarised standard stars effectively fall within 1$\sigma$ of the published value, with one additional within 2$\sigma$.
}

\begin{table*}
   \caption{Polarimetric results. 
    {\bfref The $Q/I$ and $U/I$ values are {\it not} corrected for the pleochroism (see text).}
   The last two columns give the 16\% and 84\% percentiles of the $P$ distributions. }
    \begin{tabular}{lclcclllc}
    \hline    \hline
    \noalign{\smallskip}
     {Object} &{Date} &{Filter} & $Q/I\;[\%]$ & $U/I\;[\%]$ & $P\;[\%]$ & $P_{16\%}\;[\%]$ & $P_{84\%}\;[\%]$ & $N_\theta^3$ \\
    \noalign{\smallskip}
    \hline
    \noalign{\smallskip}
      DENIS J2200 & 01/08/05 & I-Bessel &$+0.08 \pm 0.09 $&$ -0.01 \pm 0.08 $ & 0 & 0 &0.137 & 3\\ 
    \noalign{\smallskip}
    \hline
    \noalign{\smallskip}
      Kelu\,1\,AB   & 02/08/05 & I-Bessel &$+0.03 \pm 0.05 $&$ +0.13 \pm 0.10 $& 0.029 & 0 & 0.201 & 2\\ 
                           & 02/08/05 &  z-Gunn  &$-0.02 \pm 0.06 $&$ +0.00 \pm 0.07 $& 0 & 0 & 0.036 & 2 \\ 
    \noalign{\smallskip}
    \hline
    \noalign{\smallskip}
      2MASS J0036 & 01/08/05 & R-Bessel & $-0.11 \pm 0.11$ & $+0.10 \pm 0.09$ & 0.065 & 0.020 & 0.222 & 3 \\ 
                                 & 01/08/05 & I-Bessel & $-0.08 \pm 0.03$ & $-0.02 \pm 0.05$ &0.077 & 0.048 & 0.107 & 3 \\ 
                                & 10/08/06 & I-Bessel &  $-0.06 \pm 0.10$ & $-0.02 \pm 0.12$ & 0 & 0 & 0.112 & 4\\ 
                               & 11/08/06 & I-Bessel & $-0.03 \pm 0.05$ & $-0.00 \pm 0.07$ & 0 & 0 & 0.054 & 4\\ 
                          &  average$^1$ & I-Bessel & $-0.06\pm0.03$ & $-0.01\pm0.04$ & 0.053 & 0.029 & 0.079 & ...\\ 
    \noalign{\smallskip}
    \hline
    \noalign{\smallskip}
      2MASS J2158 & 01/08/05  & I-Bessel &$-0.01 \pm 0.22 $&$-0.03 \pm 0.26 $& 0 & 0 & 0.057 & 3\\ 
                                & 02/08/05  &  z-Gunn  &$-0.08 \pm 0.11$ & $-0.19 \pm 0.19$ & 0 & 0 & 0.319 & 3\\ 
    \noalign{\smallskip}
    \hline
    \noalign{\smallskip}
      LHS\,102BC       & 01/08/05 & I-Bessel &$+0.38 \pm 0.06 $ & $-0.03 \pm 0.12$ & 0.376 & 0.315 & 0.435 & 2 \\ 
                                    & 02/08/05 & I-Bessel &$+0.23 \pm 0.15 $ & $-0.08 \pm 0.22$ & 0.204 & 0.057 & 0.358 & 2 \\ 
                                    & 02/08/05 & z-Gunn  &$-0.5   \pm 0.7   $&$-0.6 \pm 0.6$& 0 & 0 & 1.209 & 3 \\ 
                                    & 28/07/06 & I-Bessel &$+0.30 \pm 0.33 $& $+0.22 \pm 0.10$ & 0.321 & 0.191 & 0.499 & 4 \\ 
                            & average$^1$ & I-Bessel &$+0.30 \pm 0.05 $& $+0.04\pm0.07$ & 0.300 & 0.246 & 0.355 & ...\\ 
    \noalign{\smallskip}
    \hline
    \noalign{\smallskip}
      2MASS J1507 & 01/08/05 &  R-Bessel & $+0.10\pm 0.15$ & $+0.04\pm 0.27$  &  0 & 0 & 0.187 & 3 \\ 
                                 & 01/08/05  & I-Bessel & $+0.02\pm 0.05$ & $-0.00\pm 0.11$  & 0 & 0 & 0.036 & 3 \\ 
    \noalign{\smallskip}
    \hline
    \noalign{\smallskip}
      DENIS J2252 & 02/08/05 & I-Bessel & $-0.07 \pm 0.09$ & $-0.11 \pm 0.06$  & 0.116 & 0.057 & 0.181 & 3 \\ 
    \noalign{\smallskip}
    \hline
    \noalign{\smallskip}
      2MASS J2244 & 02/08/05 & z-Gunn &$+0.42 \pm 0.84 $&$-0.26 \pm 0.20 $& 0 & 0 & 0.749 & 2 \\ 
                                 &                  &                &$+0.32 \pm 0.46$&$-0.07 \pm 0.15 $& 0$^2$  & 0 & 0.562 & 2\\ 
                               & 19/08/06 & I-Bessel & $+0.75 \pm 1.46$ & $-0.52 \pm  1.08$ & 0 & 0 & 1.581 & 3 \\ 
                                 &                  &                &$+0.04 \pm 1.1 $&$-0.28 \pm 0.61 $& 0$^2$ & 0 & 0.507 & 3 \\ 
                                & 21/08/06 & I-Bessel & $-0.72 \pm 0.88$ & $+0.82 \pm 0.87$ & 0 & 0 & 1.684 & 3 \\ 
                                 &                  &                 &$-0.54 \pm 1.0 $&$+0.04 \pm 1.1 $& 0$^2$ & 0 & 0.967 & 3 \\ 
                                  & 22/08/06 & I-Bessel &$-0.39 \pm 1.4 $&$+1.0 \pm 1.2 $& 0$^2$ & 0 & 1.828 & 3 \\ 
                    & average$^{1,2}$ & I-Bessel &$-0.30 \pm 0.17 $&$+0.25 \pm 0.38 $& 0.343$^2$ & 0.157 & 0.549 & ... \\ 
    \noalign{\smallskip}
    \hline
    \noalign{\smallskip}
      $\epsilon$ Indi Bab     & 01/08/05 & I-Bessel &  $+0.03 \pm 0.13$ & $+0.05 \pm 0.13$ & 0.055 & 0.000 & 0.111 & 3 \\ 
    \noalign{\smallskip}
    \hline
   \end{tabular}
  \begin{list}{}{}
  \item[$^1$]  Average based on the three $I$-Bessel measurements, the errors are propagated from the individual measurement uncertainties.
  \item[$^2$]  PSF-photometry.   {\bfref For 2MASS\,J2244, only the three PSF-photometry results are considered in the average.}
  \item[$^3$] Number ($N_\theta$) of exposures per retarder angle. 
  \end{list}
  \label{tab:Results}
\end{table*}

{\bfref The situation with non-polarised standard is not so clear: half the $Q/I$ and $U/I$ measurements are above 0.13\% (in absolute value), while the other half satisfactorily  lies within 0.07\% of a null measurement.
We find positive or negative measurements, as expected from random noise.
The polarisation vector orientation similarly does not exhibit any preferred direction. }
{For one non-polarised standard star only, on August 1, 2005, we took the 3-image-per-retarder-angle observing sequence also used for the science targets, and find a large polarimetric error {\bfref of about 0.1\%} (see Table\,\ref{polstd}). We could not find an obvious explanation for the large error.
\bfref Finally, the polarisation angle of NGC~2024 oscillates between $45\deg$ and $135\deg$, which is due the discontinuity of the polarisation angle at $Q/I=0$.
This reflects a somewhat poor measurement of that parameter, while the polarisation degree falls within 0.05\% of the published values of \citet{Fossa07}. }

{\bfref All this leads us to believe that our minimal uncertainties probably lie around 0.1\% rather than 0.05\%, at least for some targets (see Fig.\,\ref{errors}), which includes photonoise (we used very short integrations for the standard stars). In particular we do not believe our non-polarised standard stars measurements to be indicative of a systematic error on the polarisation degree or the polarisation angle.
This 0.07\% uncertainty is close or below the uncertainties on the measurements of our science targets, and is probably a minor contributor compared to the targets' photonoise.}

\section{Results} \label{results}

\subsection{New polarimetric measurements}

We report our measurements in Table\,\ref{tab:Results}.
{All measurements of $Q/I$, $U/I$ and $P_L$ {appear to be within $3\sigma$ of a null polarisation, with a vast majority within $2\sigma$.
The only exception is the $Q/I=-0.38\pm 0.06$\% of LHS102BC on August 1, 2005.}
Because this measurement has an uncertainty much smaller than usual for its SNR of 186 (see Fig.\,\ref{errors}), this uncertainty could be (statistically) underestimated, and the detection may be insignificant.
More convincingly, the three $Q/I$ measurements display a small dispersion of 0.06\% for an average of $\langle Q/I \,\rangle=-0.31\%$.
Two other targets, 2MASS~J0036 and 2MASS~J2244, have three $I$-band measurements, which are all compatible within 2\,$\sigma$. 
This prompts us to also calculate the average of the Qs and Us. Again we use the standard deviation to determine the uncertainty. We also fail to detect significant polarisation on the averaged values of 2MASS~J0036 and 2MASS~J2244.
Therefore we find only target, LHS102BC, to show a significant polarisation.
\bfref We measure a polarisation angle of $-6.6\pm9.3\deg$.
}

Taking these results at face value {leads to a fraction of polarised ultra-cool dwarfs detected in the optical smaller in value than reported by M02 (about 50\%) and ZO05 ($29\pm 9\%$).} 
Our full sample of nine targets leads to a fraction of $11^{+18}_{-4}\%$ (1-$\sigma$), and smaller than 39\% at 95\% of confidence level (C.L.), but some of our measurements have large uncertainties.
If we restrict the sample to those targets with polarisation errors smaller than 0.2\%, we find eight targets observed usually in the $I$ band (but also once in the $z$ band and twice in the $R$ band), leading to an upper-limit of 28\% (95\% C.L.) of ultra-cool dwarfs with a polarisation degree larger than 0.5\%, in agreement with previous works (M02,ZO05).

We note that we treat binaries as if they were single objects.
This is essentially true for $\epsilon$ Indi BC and DENIS\,J2252, which have a small brightness ratio in the $I$ band, so that only the warmer primary contributes to the flux.
The cases of DENIS\,J2200 {and of LHS102BC are} different, being nearly-equal-brightness binaries. {\bfref The contribution of each nearly equal-mass components to the observed linear polarisation of J2200 and LHS102BC is unknown. Both components may contribute with a similar amount provided the fact they share very similar atmospheres, parallel rotation axis and nearly identical rotation rates. If only one of the objects of the pairs is polarised,} our sensitivity to the polarised light of the latter would be decreased by a factor of two.

\subsection{Comparison with previous surveys}

As mentioned in Section\,\ref{selection}, we preferentially selected targets with previous polarimetric measurements, in order to search for variability. Seven of our targets have previous $I$ data, with 2MASS\,J0036 also having previous $R$ data. {\bfref We compare our measurements in Table\,\ref{tab:Comp} (see also Fig\,\ref{model}). In that table we use the best single-epoch measurement of ZO05 for 2MASS\,J0036 in $R$.}

Our polarisation degrees along $Q$ and $U$ fall within $3\sigma$ of those of M02, also obtained with FORS1, {with the exception of $P_Q(\rm LHS102BC)$}. 
The observing set-up of the latter study is different, with 16\,retarder angles used, and only one image per angle. This allows a different error determination and a better removal of the instrumental polarisation and other systematics. The total integration time is similar (within $\approx 50\%$) but the seeing of our images is significantly worse. 
{The $P_Q$s of LHS102BC differ by 5$\sigma$.}

Compared to ZO05 we tend to find smaller polarisation degrees (six times out of seven). 
Both observing set-ups are strictly identical, but the instruments differ.
{The repetition of the pattern is unexpected.}
Variability could be proposed to explain a fraction of the observed changes, but cannot possibly explain all {six measurements.
Some formally-significant differences in $P_Q$ or $P_U$ are certainly due to an underestimated uncertainty when the errors on $P_Q$ and $P_U$ widely differ in ZO05 (see also our Fig.\,\ref{errors}).
Eventually, only two measurements show robust discrepancies: $P_U$ for 2MASS\,J1507 and $P_Q$ for 2MASS\,J2244.

{\bfref Apart from intrinsic variability, differences in our measurements could be due to the different bandpasses, with FORS1 $I$ filter 72\,nm bluer than CAFOS's. 
However, \citet{Sengu01} and \citet{Sengu05} both predict an increasing polarisation degree with wavelength over 0.6--1\,$\mu$m for a wide range of dust grain size (results are presented for 0.1--1\,$\mu$m). {\bfref \citet{Sengu03} presents similar conclusions over 0.7--0.9\,$\mu$m for particle sizes between 0.2\,$\mu$m and 30\,$\mu$m.}
Under their hypotheses, the difference in effective wavelength is not the right explanation to the different results.
}

Among all three studies, although we have a sample of well-measured targets with multiple epochs relatively small, we do find three brown dwarfs with seemingly variable polarisation degree. 
This result needs to be confirmed with a more homogeneous data set, but offers a promising prospect for the study of the dust cloud dynamics.
}

%%%%%%%%%%%%%% COMPARISON
\begin{table}
   \caption{Comparison of our polarisation degrees with published measurements, in percent (1-$\sigma$ interval or error).
   For each target and filter, the second line gives the 2-$\sigma$ interval, correcting the published values following \citet{Simmo85}.
   }
    \begin{tabular}{lcccc}
    \hline    \hline
    \noalign{\smallskip}
     {Object} & {F} & this paper & ZO05 & M02  \\
    \noalign{\smallskip}
    \hline
    \noalign{\smallskip}
      Kelu\,1\,AB      & $I$& 0--0.20 &  $0.80\pm 0.27$ & ... \\  
                                 &   & 0--0.30 & 0.36--1.22 &  ... \\ 
      2MASS J0036 &$R$& 0.02--0.22 & $0.61\pm0.20$ & ... \\  
                                 &     & 0--0.32 & 0.39--0.81  & ...  \\ 
                                 &$I$ & 0.03--0.08$^1$ & $0.05\pm0.06$ & $0.199\pm0.028$   \\ 
                                 &     & 0--0.11$^1$ & 0--0.13 & 0.141--0.253   \\ 
      2MASS J2158 & $I$& 0--0.06 & $1.38\pm 0.35$ & ... \\  
                                 &    & 0--0.23 & 0.80--1.91 & ...  \\
      LHS\,102BC       & $I$& 0.246--0.355$^1$ & ... & $0.105\pm0.036$\\
                                    &   & 0.191--0.409$^1$ & ... &0.032--0.171 \\ 
      2MASS J1507 &$I$& 0--0.04  & $1.36\pm 0.30$ & ... \\ 
                                 &   & 0--0.09 & 1.01--1.69 & ...  \\ 

      DENIS J2252 & $I$& 0.06--0.18  & $0.62\pm0.16$ & ...  \\ 
                                &     & 0--0.25 & 0.38--0.84 & ...  \\

      2MASS J2244 & $I$& 0.16--0.55$^{1,2}$  & $2.45\pm0.47$ & ...  \\ 
                                 &    & 0--0.76 & 1.90--2.98 & ... \\ 
    \noalign{\smallskip}
    \hline
    \noalign{\smallskip}
                           &       & $\lambda_c\pm$FWHM & $\lambda_c\pm$FWHM& $\lambda_c\pm$FWHM \\
                           &       & [nm] & [nm] & [nm]  \\
    \hline
    \noalign{\smallskip}
       Filter           &$R$& $657\pm150$ & $641\pm158$ & ...  \\
       Filter           &$I$& $768\pm138$ & $850\pm150$ & $768\pm138$ \\
    \noalign{\smallskip}
    \hline
   \end{tabular}
  \begin{tabular}{p{8.5cm}}
  $^1$  Averaged value; $^2$  PSF-photometry; $^3$ Central wavelength and full-widths at half-maximum of the filter profiles, in nm.
   \end{tabular}
    \label{tab:Comp}
  \end{table}
  
\subsection{Compilation of existing data} \label{existing}

In order to increase the size of the sample, we collect measurements from the work of M02 and ZO05.
Again we restrict the sample to well-measured L-dwarf targets, with polarimetric uncertainties smaller than 0.2\%.
For consistency we only consider $I$-band data.
We count multiple non-detections  (2MASS\,J0036 in our data and those of ZO05) {and multiple detections (LHS102BC in ours and M02's)} as one measurement.
In contrast, we count conflicting results (again 2MASS\,J0036, in M02) twice, which reflects the low significance of each individual observations, {or possible intrinsic variations.}
The fraction of polarised dwarfs which we derive really is the probability to detect polarised light using (various) non-ideal instruments, while the true fraction is necessarily higher (even assuming the measurements and uncertainties are correct).

We have {\bfref seven} such objects in our sample, {\bfref 17} in ZO05 and all eight targets of M02.
We have one polarised target at the 3-$\sigma$ level.
M02 has five (including LHS102BC, already counted). 
ZO05 have five targets showing significant polarisation degrees. 
{ZO05 do not correct for the intrinsic bias in the polarisation degree.
However, their targets showing significant polarisation degree also display a significant polarisation in a single Stokes parameter (in $U/I$), hence the detections are not due to that bias.
}
The total is nine objects among 30. 
The fraction of polarised L dwarfs is therefore $30^{+10}_{-7}\%$ (1-$\sigma$ errors). 

Restricting further the sample to targets with 0.1\% uncertainties and defining polarised objects as those with polarisation degrees measured to be larger 0.3\%, 
we obtain five targets in our sample, 
all eight of M02 and 
five in ZO05 including again 2MASS\,J0036, 
for a total of 19. %16. 
{Only LHS102BC exhibit such a large polarisation degree (this paper).
The statistical probability, given our errors, that it has a polarisation degree smaller than 0.3\% is about 50\%, so it counts as $1/2$ as polarised and $3/2$ as unpolarised (M02).}
The upper-limit on the fraction of polarised ultra-cool dwarfs is now 16\% (95\% confidence level). 

\section{Interpretation}

{\bfref We have outlined in the section\,\ref{intro} an ambitious polarimetric study of brown dwarfs.
As described above, given the small signal and the diversity of observing set-up  (with M02) and of instruments (with ZO02), our results of polarimetric variability require additional data for confirmation.
We can however attempt to compare our results with the models of \citet{Sengu05} and with those of the atmospheric variability surveys.
}

\subsection{Comparison with  \citet{Sengu01} models, and correlation with flattening}

\begin{figure}[t]
\includegraphics[width=.5\textwidth,height=.5\textwidth]{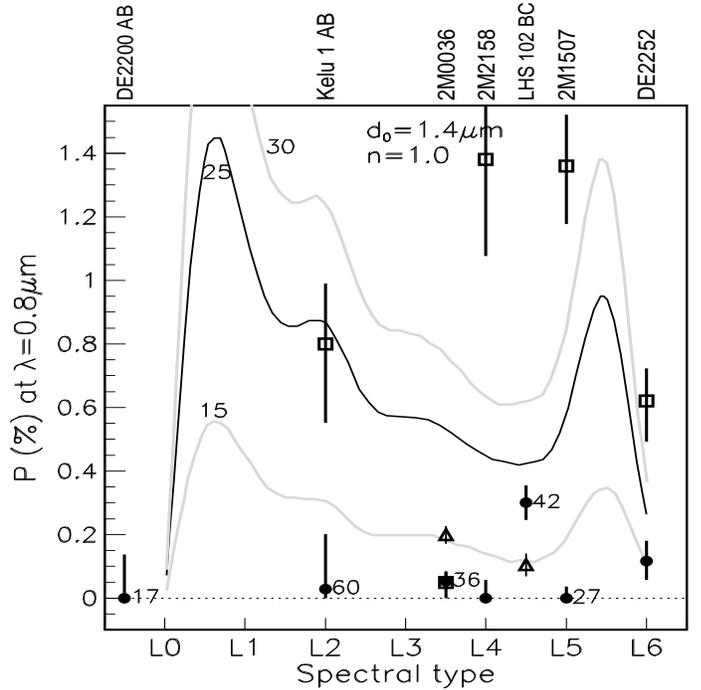}
   \caption{Polarisation degrees as a function of spectral type for our sample \bfref (filled circles), ZO05 (empty squares: published values; the error bars are corrected for the polarisation bias) and M02 (empty triangles), for the same targets. We superimpose the \citet{Sengu05} predictions for various rotational velocities (starting from the bottom: 15, 25, 30\,km/s).
   We {\bfref also} report $v\sin i$ measurements {\bfref of the sources} (in km/s) when available.
   }
   \label{model}
\end{figure}

Despite our small sample size, we can compare the non-detection of polarisation for the five ultra-cool dwarfs with $v\sin i$ measurements (see Table\,\ref{tab:Objects} and Fig.\ref{model}).
Four dwarfs have larger-than-average (projected) rotational velocities, with \mbox{$v\sin i>25\,$km/s}.
{The inclination $i$ is unknown, so how does the polarisation depend on it? At a given rotational velocity, the projected} flattening of the dwarf decreases when $\sin i$ decreases, i.e. when seen from the pole (in the extreme case of $\sin i=0$), and hence the polarisation degree {decreases as well, linearly with $\sin i$.
However, at a given projected rotational velocity}, the true rotational velocity would increase by $(\sin i)^{-1}$, and the flattening would increase: 
{Following the Eqs. 1 and 2 of \citet{Sengu01}, for most relevant eccentricities, i.e. for $0.04<e<0.9$\footnote{
{\bfref The flattening $f$($\ll$1), is related to the eccentricity $e$ by the relation: $e^2=1-(1-f)^2$.
Jupiter has a flattening of 0.065 corresponding to} an observed eccentricity of 0.35. A 50-M$_{\rm Jup}$, 0.7-R$_{\rm Jup}$ brown dwarf with the Jupiter rotational period of 9.9\,h would have a predicted eccentricity of 0.04.},
the difference of the equatorial and polar radii is roughly proportional to $(\sin i)^{-2}$, and therefore the polarisation degree is nearly independent of $\sin i$.}

For a 1.4-$\mu$m grain size, Figure\,3 of \citet{Sengu05} predicts a polarisation degree larger than 0.5\% for  a polytropic index $n=1$ and rotation velocities of 25\,km/s and higher, with a minimum at L4.5 and 0.5\% polarisation degree.
Our data strongly exclude these predictions, so that under the hypothesis of \citet{Sengu05}, either the typical grain size is larger than 1.4\,$\mu$m or the polytropic index is (unexpectedly) much larger.
Alternatively, some of those hypotheses {\bfref may not be applicable}: 
a thicker cloud deck would increase the multi-scattering and reduce the net polarisation.
{We also stress that those predictions assume that the scatterers are spherically symmetric \citep{Simmo82}.} 
The polarisation due to scattering on non-spherical grains is {\bfref barely understood and extremely difficult} to predict. Non-spherical grains could induce much smaller polarisation degrees.

{\bfref Finally, the polarisation predictions are based on atmospheric temperature-pressure profiles, as well as dust content models, such as those of Marley (1999, priv.com.), \citet{Burro01} and \citet{Coope03}.
Those models are found to reproduce reasonably well the spectral energy distribution (SED) of L and T dwarfs, but the resulting polarisation prediction may be more sensitive to the input parameters than the SED.

Because it is so difficult to predict the polarisation degree, it is equally difficult to predict the fraction of brown dwarfs with a polarisation degree larger than some given value, which is what we measure.
Instead we could compare the distribution of polarisation degrees, or its upper-limits, with spectral type.  \citet{Sengu05} predicts a larger polarisation degree between L0 and L1 and between L5 and L6.
We do not detect polarisation for our targets of those spectral types: DENIS~J2200, DENIS~J2252, or 2MASS~J1507, which has a projected rotational velocity of 27\,km/s, a median value for brown dwarfs.
This does not come in support of the hypotheses made by \citet{Sengu05}, but clearly our statistics is very limited.

\subsection{Comparison with atmospheric variability studies}

Regarding the impact of cloud cover patchiness, our results should be easier to compare with those of the spectro/photometric variability studies. Indeed, the linear polarisation degree is expected to increase for larger physical scale of the surface heterogeneities, just as the variability signal due to the brown dwarf rotation or to the dynamical variations in the cloud cover. On the other hand, as noted by Sengupta (priv.com.) in ZO05, atmospheric variability may require thick clouds to be noticeable, and the clouds' thickness would result in multiple scattering and a reduced polarisation degree.

On the observational side the matters are even more complicated. For the polarimetric studies we suffer from a theory limitation: we cannot relate a (detectable) polarisation degree to a physical size,
while variability studies suffer from a lack of reproducibility: depending on the survey, the fraction of variable objects vary from 10\% to 50\%, as the different instrument sensitivity and stability, and classification methods, vary widely \citep{CBJ05,Goldm05,Koen05}. Moreover, very few objects or even none (again depending on the definitions) have shown repeated signs of similar variability, or a stable periodic variability, as expected for rotationally-driven variations.

It is therefore futile to compare the fraction of polarised brown dwarfs and of variable ones. 
Among our targets several have monitored for photometric and/or spectroscopic variability:
\begin{itemize}
\item Kelu\,1\,AB probably has the best case for variability \citep{Clark02,Clark03,Cabal06}, and we do not find it to be polarised, above the 0.3\% level (ZO05 do).
\item 2MASS\,J0036 has been detected as variable by \citet{Berge05} in its radio emission, by \citet{Maiti07} in the $R$ and $I$ bands (not simultaneously), not by \citet{Gelin02} and \citet{Cabal03}, and we do not find it to be polarised above the 0.13\% level (M02 and ZO05, {\bfref in $R$ only,} do).
\item 2MASS\,J1507 has shown some limited variability, over a few hours (Goldman et al, {\it in prep.}), but not by \citet{Koen03}, and we do not find it to be polarised above the 0.1\% level (ZO05 do). 
\item \citet{Moral06} found some evidence for variability  in their Spitzer/IRAC monitoring of 2MASS\,J2244 at 4.5\,$\mu$m. We do not detect polarisation above the 0.76\% level, while ZO05 do.
\end{itemize}

This compilation reveals a complicated and possibly contradictory picture. Apart from the possible and general over-confidence of authors in their results, we may conclude that the atmospheres of brown dwarfs could be very dynamic and variable, even at large physical scales.
}

\section{Conclusions} \label{conclusion}

We have obtained FORS1 polarisation measurements in the $R$, $I$ and/or $z$ band for nine field ultra-cool dwarfs and binaries with spectral types of the primary from M9 to T1.
{We detect a significant polarisation for LHS102BC, with \mbox{$P_L=(0.31\pm0.06)\%$.}
We failed to detect polarisation at the 3-$\sigma$ level in the other targets,}
{setting an upper-limit to the fraction of ultra-cool dwarfs with polarisation larger than 0.5\% to 28\% (95\% confidence level)}.
We tend to find smaller polarisation degrees than \citet{ZOsor05}, and larger uncertainties than \citet{Menar02}, probably due in part to worse observing conditions and different observing strategy.
Combining all three data sets, we set an upper-limit of 16\% (95\%C.L.) to the fraction of polarised ultra-cool dwarfs with polarisation degrees larger than 0.3\%.

{\bfref 
Our small sample does not reproduce the predictions of \citet{Sengu05} regarding the flattening-induced polarisation of brown dwarfs of known rotational velocities.
Possible explanations include the effects of non-spheroidal grains, of multi-scattering, or of the underlying atmospheric models, be it the dust distribution or  the grain size distribution.
Regarding the polarisation induced by heterogeneous cloud coverage, we fail to find a correlation between the polarisation measurements and the results of atmospheric variability studies.
The discordant results of both polarimetric or variability studies may be due to a complex dynamics of the ultra-cool dwarf atmospheres.
}

\begin{acknowledgements}
       We thank F.\,M\'enard for his useful explanations, S.\,Sengupta for his comments, the anonymous referee for his/her wise suggestions, and the ESO staff for its support and for conducting the observations.
       This Research has made use of the {\sc Simbad} database, operated at C.D.S., Strasbourg, France,
       and of the M, L, and T dwarf compendium housed at {\tt DwarfArchives.org} and maintained by C.\,Gelino, D.\,Kirkpatrick, and A.\,Burgasser.
 \end{acknowledgements}

\bibliographystyle{aa}
\bibliography{mybib.bib}

\end{document}